\documentclass{aa}
\usepackage{graphicx}
\usepackage{natbib}

\begin{document}
   \title{Sulphur abundances in metal-poor stars\,\thanks{Data from the UVES Paranal Observatory Project (ESO DDT Program ID 266.D-5655) was used for this research.}}


   \author{A.J. Korn
          \and
          N. Ryde
          }
   \offprints{Andreas Korn}

   \institute{Department of Astronomy and Space Physics, Box 515, SE-75 120 Uppsala, Sweden\\
              \email{korn@astro.uu.se, ryde@astro.uu.se}
 }

   \date{Received; accepted }

   \abstract{We investigate the debated ``sulphur discrepancy'' found among metal-poor stars of the Galactic halo with [Fe/H]\,$<$\,$-$2. This discrepancy stems in part from the use of two different sets of sulphur lines, the very weak triplet at 8694\,-\,95\,\AA\ and the stronger triplet lines at 9212\,-\,9238\,\AA. For three representative cases of metal-poor dwarf, turnoff and subgiant stars, we argue that the abundances from the $\lambda\lambda$8694\,-\,95 lines have been overestimated which has led to a continually rising trend of [S/Fe] as metallicity decreases. Given that the near-IR region is subject to CCD fringing, these weak lines become excessively difficult to measure accurately in the metallicity regime of [Fe/H]\,$<$\,$-$2. Based on homogeneously determined spectroscopic stellar parameters, we also present updated [S/Fe] ratios from the $\lambda\lambda$9212\,-\,9238 lines which suggest a plateau-like behaviour similar to that seen for other alpha elements.

   \keywords{Stars: abundances, Stars: atmospheres,  Stars: Population II,
  Galaxy: abundances, Galaxy: evolution, Galaxy: halo
               }
   }

   \maketitle
%

\section{Introduction}
 Damped Ly$\alpha$ systems  are thought to be embryonic galaxies at high redshifts.
  Sulphur is an important probe of the chemical enrichment of these
  systems and can in principle be used to study star-formation histories at cosmological distances \citep{PEN:04}.
 Sulphur is not depleted onto dust grains, which is why it can be used in studies of this kind.
However, to use an element as a cosmochemical probe, its nucleosynthesis history has to be mapped out and understood. For observational reasons, this is best done in our own Galaxy. Sulphur has traditionally been viewed as
  an $\alpha$-element with a behaviour as a function of metallicity similar to that of Mg, Si and Ca,
 their production site assumed to be Supernovae (SNe) Type II. The abundance ratio of [S/Fe]\footnote{The customary bracket notation is defined as follows:
$\mathrm{[A/B]\equiv \log(A/B)_\star-\log(A/B)_\odot}$} vs. [Fe/H] in halo stars is, therefore, expected to only be
 affected by these events and would show a constant level of [S/Fe]\,$\approx$\,0.3 for [Fe/H]$<-1$, i.e.\ below a metallicity where SNe Type Ia could contribute significant amounts of metals to the stellar inventory of the Galaxy.

 This traditional view of the abundance trend in the halo phase of the Milky Way has been questioned by \citet{israel} and \citet{takeda}
 who, based on data from large telescopes, found a [S/Fe] trend linearly increasing
 with decreasing metallicity. In order to explain this overproduction of sulphur at low metallicities, they suggested an extra component for the production of sulphur at low metallicities, such as hypernovae.
 Later, however,  \citet{PEN:04}, \citet{ryde:04_S} and Ryde \& Lambert (2005, {\sl in press}) contradicted this finding, and argued for the traditional plateau-like behaviour
 in the halo. The reason for this difference in point of view is partly that these authors used a different, arguably more reliable, diagnostic and partly that they used iron abundances based on Fe\,{\sc ii} lines instead of neutral iron. Recently, \citet{Takada_pasj} and Takeda et al.
 (PASJ, {\sl in press}) partly corroborate the plateau behaviour and therefore the $\alpha$-element nature of sulphur. But they also find a discrepancy between the sulphur abundance derived from different sulphur multiplets.

Analysing the sulfur abundance of metal-poor stars is problematic due to the lack of suitable atomic lines. Diagnostic
S\,{\sc i} lines lie in the near-infrared, with lines of
high-excitation occurring at 8693.2\footnote{Like in all the other mentioned studies, this line is disregarded, as it is the weakest of the multiplet.}, 8694.0 and 8694.6\,\AA\ and 9212.9, 9228.1 and 9237.5\,\AA.
 \citet{PEN:04} and \citet{ryde:04_S} used the lines lying at 9212\,-\,9238\,\AA\ (multiplet 1) instead of the much weaker lines at 8694\,-\,95\,\AA\ (multiplet 6) which were used by \citet{israel} and \citet{takeda}. A drawback with the later lines
 is their weakness in halo stars, making their analysis very difficult and subject to large, and sometimes systematic,
 uncertainties. A drawback with the lines at 9212\,-\,9238\,\AA\ is the forest of telluric lines that affects this wavelength region. \citet{PEN:04} and \citet{ryde:04_S}  showed, however, that it is possible to process the observational data in order to reduce this effect, with very good results.

In this \emph{Research Note} we revisit the discrepancy of the sulphur abundances
in metal-poor stars. We argue that the ability to accurately measure the weak lines around 8693\,\AA\ has been systematically overestimated. At least for relatively unevolved stars (main-sequence to subgiant stars), we find the observations to be compatible with a plateau-like low sulphur abundance as derived from the $\lambda\lambda$\,9212\,-\,9238 lines which are typically ten times stronger than the lines at 8694\,-\,95\,\AA.

\begin{table}
      \caption[]{Line parameters}
         \label{line_parameters}
     $$
         \begin{array}{rrrrr}
            \hline
            \noalign{\smallskip}
        \mathrm{Wavelength }     & \textrm{E$_\mathrm{exc}$} & \textrm{$\log\,gf$}  \\ 
            \noalign{\smallskip}
           \textrm{(\AA )} & \textrm{(eV)} &   \mathrm { (cgs)} \\
           \noalign{\smallskip}
              \hline
           \noalign{\smallskip}
            \multicolumn{3}{l}{\textrm {S \sc{i}~} {\rm lines}}\\
            \noalign{\smallskip}

 8693.958  &   7.870  &   -0.51^{\mathrm{a}}   &   \\ 
  8694.641  &   7.870  &    0.08^{\mathrm{a}}  &  \\ 
  9212.863  &   6.525  &    0.43^{\mathrm{b}}   &   \\ 
  9228.093  &   6.525  &    0.25^{\mathrm{b}}  &   \\ 
  9237.538  &   6.525  &    0.03^{\mathrm{b}}  &    \\ 
          \hline

         \end{array}
     $$
\begin{list}{}{}
\item[$^{\mathrm{a}}$] Line data taken from \citet{takeda}
\item[$^{\mathrm{b}}$] Line data taken from the {\sc nist}
database

\end{list}
   \end{table}

\section{VLT UVES data}
\citet{israel} have three stars with [Fe/H]$<-1.75$ for which they have measured a sulphur abundance from the $\lambda\lambda$\,8694\,-\,95 lines,
namely HD\,2665, HD\,2796 and HD\,19445. \citet{takeda} present corresponding sulphur abundances for two stars: HD\,165195 and HD\,84937.
Furthermore, \citet{Takada_pasj} present new sulphur abundances based on the $\lambda\lambda$\,8694\,-\,95 lines for five stars, namely
HD\,140283, HD\,187111, HD\,216143, HD\,221170, and BD\,+371458. For HD\,140283 and HD\,216143 they find significantly higher sulphur abundances
from the $\lambda\lambda$\,8694\,-\,95 lines compared to those from the $\lambda\lambda$\,9212\,-\,38 lines. For a total of five of these stars, sulphur abundances from both sets of lines are reported in the literature (cf.~Table \ref{stars}).
We collected high-resolution ($R$\,=\,80\,000) data for two of these stars (HD\,84937 and HD\,140283), freely available from the UVES Paranal Observatory Project (Bagnulo et al. 2003). Spectra of the other stars mentioned above are not available in the UVES POP database. For a third star (HD\,19445), we have high-quality data available to us \citep{korn:03} which was taken with the fibre-fed spectrograph FOCES \citep{pfeiffer:98} at $R$\,=\,60\,000. This is a small, albeit comprehensive, data set to check upon the observational reality of this discrepancy.

Using the MAFAGS suite of codes (Gehren 1975a,\,b and subsequent updates, see \citet{fuhrmann:97}), we have computed synthetic spectra for the $\lambda\lambda$\,8694\,-\,95 lines for the three dwarf and subgiant stars listed in Table~\ref{stars}, for two values of the sulphur abundance as given in column 9 and 10. The stellar parameters were taken from the respective analyses, as quoted in Table~\ref{stars}. External broadening (macroturbulence, instrumental broadening and stellar rotation) was estimated from a nearby Fe\,{\sc i} line at 8688.63\,\AA\ and modelled in the Gaussian approximation. This iron line was also used to correct for the stellar radial velocity.

For all three stars, the two synthetic spectra are compared with observations in Figure~1.

   \begin{table*}
      \caption[]{Stars from the literature that show a large discrepancy between the two sulphur diagnostics}
         \label{stars}
     $$
        \begin{array}{llcccccccc}
            \hline
            \noalign{\smallskip}
        \mathrm{Star}    & \mathrm{Spectral} & \mathrm{V} &  T_{\mathrm{eff}}  & \log\,g & \mathrm{[Fe/H]} & \xi_\mathrm{micro} & \xi_\mathrm{macro} &  \multicolumn{2}{c}{\log\,\varepsilon{\rm(S)}^{\rm b}} \\
            & \mathrm{type}^\mathrm{a} & \mathrm{[mag]} & \mathrm{[K]}& \mathrm { (cgs)} & &\mathrm{[km\,\, s^{-1}}] &\mathrm{[km\,\, s^{-1}}] & \textrm{$\lambda\lambda$\,9212\,-\,38} & \textrm{$\lambda\lambda$\,8694\,-\,95} \\
            \noalign{\smallskip}
          &  & &   &  & \pm0.15 & \pm0.5 & \pm0.5 &  \pm0.15 & \pm\sim0.15\\
            \noalign{\smallskip}

            \hline
           \noalign{\smallskip}
            \multicolumn{5}{l}{\textrm {Giants}}\\
            \noalign{\smallskip}
           \textrm {HD~2665} & \textrm {G5III} & 7.8 & 4990\pm 
           70 & 2.50\pm 0.20  & -1.74^1/-2.00^2 & 1.5&  5.5  & 5.60^1 & 5.89^2 \\
           \textrm {HD~216143} & \textrm {G5}  & 7.8 & 4540\pm 100 & 1.78\pm0.15 & -2.15^4 & 2.1 & - & 5.48^4 & 5.72^4 \\

           \hline
            \noalign{\smallskip}
            \multicolumn{5}{l}{\textrm {Dwarfs and subgiants}}\\
            \noalign{\smallskip}
          \textrm {HD~19445}  & \textrm {sdG5}  & 8.0 & 5810\pm 150 & 4.46\pm0.20 & -1.90^1/-1.88^2  & 1.5 & 4.5  & 5.50^1 & 5.97^2 \\
           \textrm {HD~84937}  & \textrm {sdF5}  & 8.3  & 6300\pm 100 & 3.97\pm0.15 & -2.06^1/-2.11^3 & 1.1  & 5.0  & 5.27^1  & 5.70^3 \\
           \textrm {HD~140283}  & \textrm {sdF3}  & 7.2  & 5830\pm 100 & 3.67\pm0.15 & -2.55^4 & 1.9  & -  & 5.18^4  & 5.53^4 \\
            \noalign{\smallskip}
            \hline
          \noalign{\smallskip}
           \multicolumn{7}{l}{\textrm {(1): \citet{ryde:04_S} using $T_{\rm eff}$, log\,$g$ and $\xi$ from (2) and (3)}}\\
          \multicolumn{4}{l}{\textrm {(2): \citet{israel}}}\\
           \multicolumn{4}{l}{\textrm {(3): \citet{takeda}}}\\
           \multicolumn{4}{l}{\textrm {(4): \citet{Takada_pasj}}}\\
            \noalign{\smallskip}
         \end{array}
     $$
\begin{list}{}{}
\item[$^{\mathrm{a}}$] From Simbad
({\tt{http://simbad.u-strasbg.fr}})
\item[$^{\mathrm{b}}$] $\log \varepsilon \mathrm{(S)}\equiv\log n_\mathrm{S}/n_\mathrm{H} + 12$; $\log$\,$\varepsilon$(S)$_\odot$\,=\,7.20 \citep{chen}
\end{list}
   \end{table*}
%

   \begin{figure*}
   \centering
   \includegraphics[angle=90,width=\textwidth]{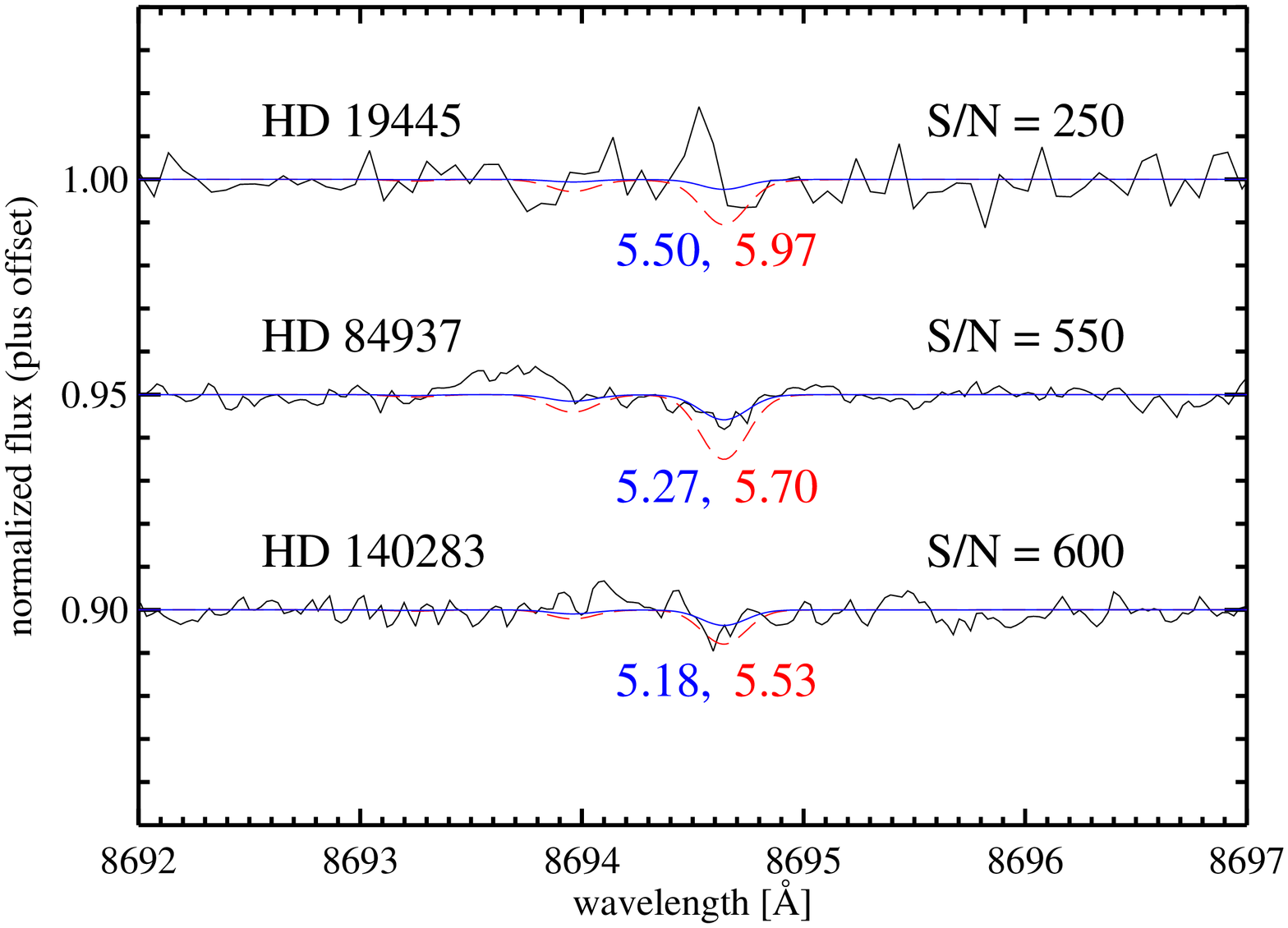}
   \caption{Spectra of the program stars in the region around the S\,{\sc i}
lines at 8694.0 and 8694.6\,\AA. The nominal signal-to-noise ratios (S/N) are high, but fringing residuals degrades the actual data quality. Overlaid are synthetic spectra for literature values of $\log \varepsilon$(S) which are indicated below the spectra (low and high values from Table~\ref{stars}, full-drawn and dashed line style, respectively). }
              \label{FigGam}%
    \end{figure*}

\section{Discussion}
As can be appreciated from inspecting Figure 1, even the strongest sulphur line at 8694.6\,\AA\ is hardly detected in any of the three spectra. We are, therefore, very doubtful that lines of this strength can be measured in this wavelength region without taking extra care of fringing. For example, the feature at 8693.5\,--\,8693.9\,\AA\ in the spectrum of HD\,84937 shows that fringing residuals are of the same amplitude as the stellar absorption line. The same holds true for the other two program stars. With equivalent widths below 1.5\,m\AA, the detections of S\,{\sc i} 8694.6\,\AA\ in the giant stars of Table~\ref{stars} are equally improbable. What can be said is that the high sulphur values (the synthetic profiles with dashed line style) tend to be ruled out at the 1 to 2\,$\sigma$ level. This is the main result from our re-analysis.

The impact of the fundamental stellar parameters deserves attention as well. As we took the three program stars from different analyses, we cannot expect the temperature, gravity and metallicity scales to match. This may well lead to an increase in scatter for the [S/Fe] ratios. For HD\,19445 \citet{israel} derived $T_{\rm eff}$ using the infrared-flux method (IRFM) yielding 5810\,K. If one, however, refers to the original compilation of IRFM temperatures of \cite{alonso:1996}, this star is tabulated at 6050\,K, a markedly hotter temperature which is in better agreement with temperatures obtained from Balmer profiles and UV fluxes (cf. \citet{korn:03}). For a metal-poor turnoff star such as HD\,84937, the microturbulence derived by \citet{takeda} is surprisingly small (we favour 1.8\,km/s) which can have an influence on [S/Fe] if iron lines of intermediate strength are use in determining [Fe/H].

To investigate the impact of stellar parameters, we have recomputed the [S/Fe] ratios we would derive from the $\lambda\lambda$\,9212\,-\,38 lines using equivalent widths from the literature and the stellar parameters presented in \citet{korn:03}. Instead of [S/Fe] of +0.20, +0.13 and +0.53 (for HD\,19445, HD\,84937 and HD\,140283, respectively), we get +0.28, +0.22 and +0.34 (cf.\ Table \ref{newanalysis}). The star-to-star scatter is thus significantly reduced by using a homogeneous set of stellar parameters.

\begin{table}
      \caption[]{Stellar parameters and re-derived sulphur abundances from the $\lambda\lambda$\,9212\,-\,9238 lines}
         \label{newanalysis}
\begin{tabular}{llcclcc}
\hline
\noalign{\smallskip}
star & $T_{\rm eff}$ & log\,$g$ & [Fe/H] & $\xi$ & log\,$\varepsilon$(S) & [S/Fe] \\
 & [K] & & & [km/s] & & \\
\noalign{\smallskip}
\hline
\noalign{\smallskip}
HD\,19445 & 6032 & 4.40 & $-$2.08 & 1.75 & 5.40 & 0.28 \\
HD\,84937 & 6346 & 4.00 & $-$2.16 & 1.80 & 5.26 & 0.22 \\
HD\,140283 & 5806 & 3.68 & $-$2.43 & 1.70 & 5.11 & 0.34 \\
\noalign{\smallskip}
\hline
\end{tabular}
\end{table}

It remains to be seen how other effects may affect these abundances. While neutral sulphur is a majority species throughout the atmospheres of cool stars, this does not guarantee that departures from local thermodynamic equilibrium do not exist (cf. Takeda et al., PASJ, {\sl in press}). Likewise, the assumptions made in connection with a one-dimensional plane-parallel temperature stratification and mixing-length theory could well affect the abundances systematically. In an very recent paper, Caffau et al. (A\&A, {\sl in press}) present sulphur abundances for two stars with [Fe/H]\,$\leq$\,$-2$ which seem to continue the sloping trend of higher-metallicity stars. As the $\lambda\lambda$\,9212\,-\,9238 lines were used, this could indicate the existence of sulphur-rich halo stars for which the scenario presented here offers no explanation. All these questions are left to future investigations.

\section{Conclusions}
We have reinvestigated the sulphur discrepancy for three of the five stars for which abundances derived from the two infrared sulphur multiplets exist in the literature, mostly by means of publicly available data (UVES POP survey for HD\,84937 and HD\,140203). In particular in the case of HD\,84937 the observations rule out sulphur abundances as high as $\log \varepsilon$(S) = 5.70. In the other two cases, the high S abundances find no particular support in the spectra, either. These high abundances seem to be subject to large systematic uncertainties in the data: even if the nominal S/N exceeds 500, fringing residuals might prevent one from proving the existence of lines as strong as 2\,m\AA. In this context it is worth mentioning that fibre-fed spectrographs are expected to deal with thinned-CCD fringing better than slit spectrographs like UVES: if the starlight and the light coming from the flatfield lamp illuminate the spectrograph in exactly the same way, then the flatfield correction will minimize the fringing problem.

From a re-analysis of the $\lambda\lambda$9212\,-\,9238 lines using the homogeneously determined spectroscopic stellar parameters of \citet{korn:03}, we find [S/Fe] to show an enhancement below [Fe/H]\,$\approx$\,$-$2 typical for $\alpha$-elements. This is consistent with the plateau found by \citet{PEN:04} and \citet{ryde:04_S}. Furthermore, the improved stellar parameters employed in this paper reduce the star-to-star scatter significantly.

This work highlights the benefits of high-quality public data sets such as the UVES POP survey. What is missing in this particular case is some form of monitoring of the night sky. This would be needed to perform a telluric correction for analysing, e.g., the 9212\,-\,9238\,\AA\ lines of sulphur. If we want to overcome differences from analysis to analysis, universally accepted benchmark cases will have to be established and dealt with. Well-studied stars like HD\,140283 are ideal targets in such an approach. This will be the only way to sharpen our view for what abundances can tell us about the evolution of the universe.

\begin{acknowledgements}
      This work was
supported in part by the Leopoldina Foundation/Germany under grant
BMBF-LPD 9901/8-87 and the Swedish Research Council. We thank Inese Ivans, David Yong and the referee for valuable comments on the original manuscript. NR thanks David Lambert for drawing his attention to the Galactic chemical evolution of sulphur.
\end{acknowledgements}


\begin{thebibliography}{}
\bibliographystyle{aa}

\bibitem[\protect\astroncite{{Alonso} et~al.}{1996}]{alonso:1996}
{Alonso}, A., {Arribas}, S., and {Martinez-Roger}, C., 1996,
\newblock {\aap} {313}, 873

\bibitem[\protect\astroncite{{Bagnulo} et~al.}{2003}]{bagnulo}Bagnulo et al., 2003, ESO Messenger 114, 10

\bibitem[\protect\astroncite{{Chen} et~al.}{2002}]{chen}
{Chen}, Y.~Q., {Nissen}, P.~E., {Zhao}, G., and {Asplund}, M., 2002,
\newblock {A\&A} {390}, 225

\bibitem[\protect\astroncite{{Fuhrmann} et~al.}{1997}]{fuhrmann:97}
{Fuhrmann}, K., {Pfeiffer}, M., {Frank}, C., {Reetz}, J., and {Gehren}, T.,
  1997,
\newblock {\aap} {323}, 909

\bibitem[\protect\astroncite{{Gehren}}{1975a}]{G75a} Gehren, T. 1975a, LTE-Sternatmosph{\"a}renmodelle (I), University of Kiel, Germany

\bibitem[\protect\astroncite{{Gehren}}{1975b}]{G75b} Gehren, T. 1975b, LTE-Sternatmosph{\"a}renmodelle (II), University Kiel, Germany

\bibitem[\protect\astroncite{{Israelian} and {Rebolo}}{2001}]{israel}
{Israelian}, G. and {Rebolo}, R., 2001,
\newblock {ApJ} {557}, L43

\bibitem[\protect\astroncite{{Korn} et~al.}{2003}]{korn:03}
{Korn}, A.~J., {Shi}, J., and {Gehren}, T., 2003,
\newblock {\aap} {407}, 691

\bibitem[\protect\astroncite{{Nissen} et~al.}{2004}]{PEN:04}
{Nissen}, P.~E., {Chen}, Y.~Q., {Asplund}, M., and {Pettini}, M., 2004,
\newblock {A\&A} {415}, 993

\bibitem[\protect\astroncite{{Pfeiffer} et~al.}{1998}]{pfeiffer:98}
{Pfeiffer}, M.~J., {Frank}, C., {Baumueller}, D., {Fuhrmann}, K., and {Gehren},
  T., 1998,
\newblock {\aaps} {130}, 381

\bibitem[\protect\astroncite{{Ryde} and {Lambert}}{2004}]{ryde:04_S}
{Ryde}, N. and {Lambert}, D.~L., 2004,
\newblock {A\&A} {415}, 559

\bibitem[\protect\astroncite{{Takada-Hidai} et~al.}{2005}]{Takada_pasj}
{Takada-Hidai}, M., {Saito}, Y., {Takeda}, Y., {Honda}, S., {Sadakane}, K.,
  {Masuda}, S., and {Izumiura}, H., 2005,
\newblock {PASJ} {57}, 347

\bibitem[\protect\astroncite{{Takada-Hidai} et~al.}{2002}]{takeda}
{Takada-Hidai}, M., {Takeda}, Y., {Sato}, S., et~al., 2002,
\newblock {ApJ} {573}, 614

\end{thebibliography}



\end{document}